\documentclass[useamsfonts]{pasj00}
\usepackage{graphicx}

\SetRunningHead{HASHIMOTO ET AL.}
	       {NEAR-INFRARED POLARIMETRY OF NGC 6334 V}
	       \begin{document}
		 
		 \title{Sub-Arcsecond Near-Infrared Images of \\ 
		   Massive Star Formation Region NGC 6334 V\thanks{Based on data collected 
		     at the Subaru Telescope, which is operated by the
	       	     National Astronomical Observatory of Japan.}}

		 \author{Jun \textsc{HASHIMOTO},\altaffilmark{1,2} Motohide \textsc{TAMURA},\altaffilmark{1,4}
		   Hiroshi \textsc{SUTO},\altaffilmark{1} Lyu \textsc{ABE},\altaffilmark{1}}
		 \author{Miki \textsc{ISHII},\altaffilmark{3} Tomoyuki \textsc{KUDO},\altaffilmark{4} 
		   Satoshi \textsc{MAYAMA}\altaffilmark{4}}

		 \altaffiltext{1}{National Astronomical Observatory, 2-21-1 Osawa, Mitaka, Tokyo 181-8588; 
		   hashmtjn@optik.mtk.nao.ac.jp}
		 \altaffiltext{2}{Department of Physics, Tokyo University of Science, 1-3, Kagurazaka, Sinjuku-ku, Tokyo 162-8601}
		 \altaffiltext{3}{Subaru Telescope, 650 North A'ohoku Place, Hilo, HI 96720}
		 \altaffiltext{4}{Department of Astronomical Science, Graduate University for Advanced Studies (Sokendai), \\ 
		   2-21-1 Osawa, Mitaka, Tokyo 181-8588}
		 
		 \maketitle
		 \begin{abstract}
		   
		   We present high spatial resolution (0$\farcs$3) polarimetric images 
		   in the $H$ and $K$ bands and direct images in the $L'$ and $M'$ bands    
		   of the NGC 6334 V infrared nebulae. The images show complex 
		   structures including the multi-shells and various knots
		   in the nebulae. The appearances and colors of the eastern and western
		   nebulae differ considerably. Our polarization images
		   also show differences between the illuminating sources of the nebulae:
		   the eastern nebula is illuminated by a deeply embedded mid-infrared
		   source, KDJ 4, and the western nebula by our newly detected near-infrared 
		   source, WN-A1. The degree of polarization
		   of the nebulae is very large, up to 70\% at $K$ and 60\% at $H$,
		   which is consistent with a single scattering of near-infrared radiation
		   from each source at the walls of the mass outflows.
		 \end{abstract}
		 
\section{Introduction}
NGC 6334 V is one of the far-infrared sources (total luminosity of $L \sim 10^{5} L_{\odot}$ ; 
Loughran et al. 1986) along the ridge of the massive star-formation regions in the NGC 6334 complex 
at a distance of 1.7 kpc (Neckel 1978). 
Fischer et al. (1982) detected 2 $\mu$m and 2.6 mm emission from H$_{2}$ 
and CO, respectively.  Near the H$_{2}$ emission region, Harvey and Wilking (1984) 
found an IR bipolar nebula. 
Simon et al. (1985) and Kraemer et al. (1999) showed that 
four IR sources (IRS V-1 detected at 3.5 $\mu$m and 12.5 $\mu$m , V-2 at 3.5 $\mu$m, 12.5 $\mu$m and 20.6 $\mu$m, 
V-3 at 3.5 $\mu$m, 12.5 $\mu$m and 20.6 $\mu$m, KDJ 4 at 12.5 $\mu$m and 20.6 $\mu$m) 
along the ridge of the nebula, which 
made it difficult to identify the nebula's source of illumination.

Several polarimetric observations of NGC 6334 V have been carried out.
Polarization studies can indicate the sources of illumination of the reflection
nebula in star formation regions. However, because of crowding of several sources, 
previous polarimetric observations of this region were unable to identify the source(s) convincingly:

1. Simon et al. (1985) measured IR aperture polarizations in the $H$ and $K$ bands of IRS V-1. 
They measured a degree of polarization at $H$ of $P_{\rm H} \sim$ 59\% 
at position angle $\theta_{\rm H} \sim 160^{\circ}$, and
$P_{\rm K}$ of $ \sim$ 46\% at $\theta_{\rm K} \sim 165^{\circ}$. 
They suggested that the large polarization of IRS V-1 was due to illumination
by IRS V-2 or V-3.

2. Wolstencroft, Scarrott \& Warren-Smith (1987) measured optical polarizations of OS1
which coincided with IRS V-2. They found $P_{\rm opt} \sim 38$ \% at
$\theta_{\rm opt} \sim 160^{\circ}$. They concluded that the illuminating source
of OS1 was IRS V-3 and that IRS V-3 was likely the illuminating source 
of the IR bipolar nebula. 

3. Nakagawa et al. (1990) measured $L'$ band polarizations of the bipolar nebula.
They found a symmetric polarization vector pattern of the bipolar nebula illuminated by 
a central star. They concluded that the source of the nebula was the $20 \mu$m peak
which was identified with IRS V-3 in Simon et al. (1985). 

4. Chrysostomou et al. (1994) measured $H$, $Kn$ and $nbL$ bands polarizations of 
the bipolar nebula. At the peaks of the eastern nebula, 
they found $P$ to be $\sim 60$ \% in the $H$ and $Kn$ bands
with $\theta \sim 160^{\circ}$. In the western nebula they found a centrosymmetric
polarization vector pattern polarized $\sim 50$ \% in the $Kn$ band, while $P$ was $\sim 20$ \% in the $H$ band. 
They concluded that the centrosymmetric polarization vector pattern in the western nebula
was due to illumination by IRS V-3. Since the polarization vector pattern of the eastern nebula was not 
centrosymmetric, they could not identify the illuminating source of the eastern nebula.

In general, the observations of distant star-formation regions require high spatial resolutions.
All the above polarimetric observations were conducted at over 1$''$  spatial 
resolutions. Considering both the large distance to NGC 6334 V and the crowding of the cluster of IR sources,
high-resolution imaging is essential to study of this IR nebula.
Thus we performed high resolution polarimetry in the $H$ and $K$ bands 
and imaging in the $H$, $K$, $L'$ and $M'$ bands.

\begin{table*}
  \begin{center}
    \caption{Near-IR, mid-IR and radio sources in NGC 6334 V}
    \begin{tabular}{cccccc} 
	\hline 
	\hline
	IR knot &Simon et al. &Kraemer et al.&Rengarajan \& Ho & RIGHT ASCENSION\footnotemark[$*$]  & DECLINATION\footnotemark[$*$] \\ 
	This paper&(1985)&(1999)&(1996)&(J2000) &(J2000) \\ 
	\hline
	EN-A1 &IRS V-1 & KDJ 1   &R E-2& 17 19 58.201 & -35 57 49.37 \\
	EN-A2 &IRS V-2 & KDJ 2   &...    & 17 19 58.066 & -35 57 50.30 \\
	EN-A3 &    ...   & ...     &...    & 17 19 58.305 & -35 57 49.53 \\
	WN-A1 &    ...   &(KDJ 3)  &...    & 17 19 57.785 & -35 57 50.81 \\
	WN-A2 &    ...   &(KDJ 3)  &...    & 17 19 57.790 & -35 57 49.84 \\
	...     &    ...   & KDJ 4   &R E-3 \footnotemark[$\ddagger$]& 17 19 57.461 \footnotemark[$\dagger$] \footnotemark[$\ddagger$]& -35 57 52.47 \footnotemark[$\dagger$] \footnotemark[$\ddagger$]\\
	\hline
	\multicolumn{6}{@{}l@{}}{\hbox to 0pt{\parbox{180mm}{\footnotesize
	      \par\noindent
	      \footnotemark[$*$] Positional errors are estimated to be less than 0$\farcs$1. 
	      \par\noindent
	      \footnotemark[$\dagger$] Absolute position is for KDJ 4.
	      \par\noindent
	      \footnotemark[$\ddagger$] Absolute position for R E-3, $17^{\rm h} 19^{\rm m} 57^{\rm s}.55,
	      -35^{\circ} 57' 51\farcs57$ (J2000), is estimated by assuming 
	      that R E-2 coincides with EN-A1 (see $\S$\ref{3.2} in the text).
	    }\hss}}
    \end{tabular}
  \end{center}
  \label{abs_pos}
\end{table*}

\section{Observations and Data reduction}\label{obs}
$H$ and $K$ polarimetry and $L'$ and $M'$ imaging of 
the NGC 6334 V IR nebula were conducted
on the Subaru telescope with CIAO (Tamura et al. 2000; 
Murakawa et al. 2004) on 2005 June 27. 
The pixel scale of CIAO was 0.0213 arcsec pixel$^{-1}$.
The details of the polarimeter have been described elsewhere 
(Tamura et al. 2003). 
Although no adaptive optics were employed in these observations,
the good and stable seeing enabled us to achieve a high spatial
resolution 0$\farcs$3 in the $H$, $K$, $L'$ and $M'$ bands. 
The polarizations were measured by stepping the half waveplate at
four angular positions (0$^{\circ}$, 45$^{\circ}$, 22.5$^{\circ}$ and  
67.5$^{\circ}$). 
We took 3 frames of 20 s $\times$ 6 co-adds integration per 
waveplate position in the $H$ and $K$ bands, 
4 frames of 0.5 s $\times$ 100 co-adds integration in the $L'$ band,
and 8 frames of 0.2 s $\times$ 100 co-adds integration in the $M'$ band.

The data were reduced in the standard manner of infrared image reduction:
subtracting a dark-frame and dividing by a flat-frame.
In addition, the data for each waveplate position 
($I_{0}, \ I_{45}, \ I_{22.5} \ \rm{and} \ I_{67.5}$)
were registered, and then combined. 
Stokes $I$, $Q$, $U$ parameters, degree of polarization ($P$), 
and position angle ($\theta$) are calculated as follows (see e.g., Tamura et al. 2003).  
\begin{eqnarray*}
   I\,&=&\, \{I_{0} + I_{45} + I_{22.5} + I_{67.5}\} \  / \ 2,\\
   Q\,&=&\,I_{0} - I_{45}, \ U\,=\,I_{22.5} - I_{67.5},\\
   P\,&=&\,\sqrt{\left(\frac{Q}{I}\right)^{2} + \left(\frac{U}{I}\right)^{2}},
   \ \ {\rm and} \ \theta \,=\, \frac{1}{2}\,{\rm arctan} \left(\frac{U}{Q}\right).
\end{eqnarray*}

\section{Results and Discussion}
\subsection{Imaging observations}
\begin{figure}[btph]
  \begin{center}
	\FigureFile(80mm,50mm){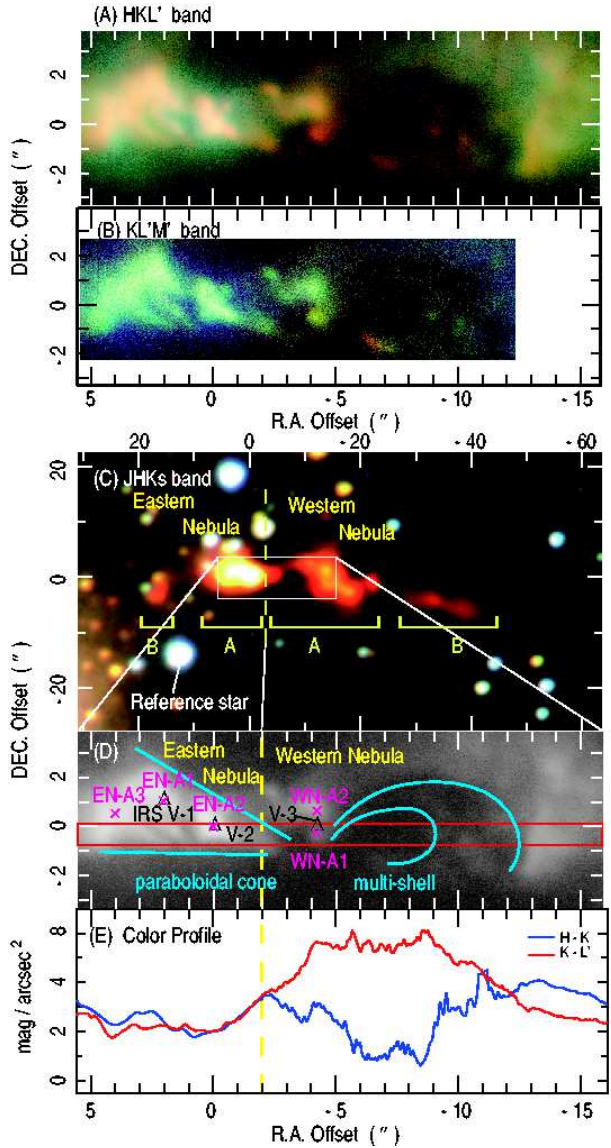}
  \end{center}
  \caption{
    (A) : The three-color composite image of the intensity 
    in the $H$, $K$ and $L'$ bands ($H$:blue, $K$:green, $L'$:red) obtained with Subaru/CIAO. 
    Only a center region of $21'' \times 7''$ of the CIAO field is shown here.
    (B) : The same as (A) but in the $K$, $L'$ and $M'$ bands ($K$:blue, $L'$:green, $M'$:red).
    (C) : The same but in the $J$, $H$ and $Ks$ band ($J$:blue, $H$:green, $Ks$:red) 
    obtained with IRSF/SIRIUS. Only a region of $\sim 100'' \times 50''$ of 
    the SIRIUS field is shown here.
    (D) : The same as (A) but this frame shows the sketch of the EN-A1 and WN-A1; 
    a paraboloidal cone and a multi-shell structure are outlined (blue line). 
    Also shown are the positions of the peak of the bright knots (pink crosses) and 
    IRS V-1 to V-3 in the 3.5 $\mu$m and 20 $\mu$m (black triangles; Simon et al. 1985). 
    Red rectangle represents a locus of the cut for a color-plofile in (E).
    (E) : A color profile of $H$ $-$ $K$ and $K$ $-$ $L'$ magnitudes averaged over the declination 
    in the red box in (D). In these all figures, the positional offsets are relative to the peak 
    at EN-A2 in the $H$ and $K$; ($0'',0''$) $= 17^{\rm h} 19^{\rm m} 58^{\rm s}.1,
    -35^{\circ} 57' 50''$ (J2000). 
  }\label{fig1}
\end{figure}

Figures \ref{fig1}(A) and (B) show 
the central region of NGC 6334 V obtained with Subaru/CIAO. 
In addition, Figure \ref{fig1}(C) shows a large scale structure of this region
taken with SIRIUS (Nagayama et al. 2003), 
a simultaneous three color infrared camera, 
on the IRSF 1.4m telescope in Sutherland, South Africa.
There are four distinct nebulae associated with NGC 6334 V. 
We refer to them as in figure \ref{fig1}(C), e.g., Eastern Nebula A, or EN-A. 

In Figures \ref{fig1}(A) and (B), the structure of the bipolar nebula is complicated. 
The shape of EN-A is a paraboloidal cone centered at $(-4'', -1'')$ 
with several bright knots, while WN-A has a multi-shell structure 
with a bright western edge. It is the first time that these fine structures 
in NGC 6334 V have been detected. 
Few previous observations have detected a multi-shell structures
(e.g., GL 2591, Tamura \& Yamashita 1992).
 
A schematic view of the structures is 
shown in Figure \ref{fig1}(D). Moreover, the colors of EN-A and WN-A  
differ systematically. Figure \ref{fig1}(E) shows that EN-A is 
bluer ($H$ $-$ $K$ $\sim$ 3 mag) than WN-A ($H$ $-$ $K$ $\sim$ 4 mag);
EN-A was detected in the $J$, $H$, $K$, $L'$ and $M'$ bands, 
while WN-A was detected in the $K$, $L'$ and $M'$ bands. 
Due to a high extinction, WN-A was detected only faintly with CIAO
in the $H$ band. 

In Figures \ref{fig1}(A) and (B), there are several bright knots which
we refer to as in Figure \ref{fig1}(D), e.g., EN-A1. Table \ref{abs_pos} summarizes 
names of these near-IR knots and their absolute positions. Each position was measured as follows:
we measured the relative position of the knots based on the IRSF/SIRIUS data,
and we used the absolute position of a reference star from 2MASS-PSC.
The position of the reference star is marked in Figure \ref{fig1} (C). 

\subsection{Identifications with previous sources}\label{3.2}
Simon et al. (1985) reported three near- and mid-IR (3.5 $\mu$m and 20 $\mu$m) sources, 
IRS V-1, V-2 and V-3.
The positions for IRS V-1 and V-2 do not exactly coincide with our new positions
for EN-A1 and A2, respectively, though they are close with each other. 
Their relative positions agree well with each other. Thus
if we assume that EN-A1 coincides with IRS V-1,
EN-A2 comes to coincide with IRS V-2. That is why we identify EN-A1 and A2 
with IRS V-1 and V-2, respectively. However, we can see no counterpart to IRS V-3.
Since IRS V-3 lies at a middle position between the newly detected WN-A1 and A2,
we suggest that our high spatial resolution observations allow IRS V-3 to be resolved into 
WN-A1 and A2. $L'$ and $M'$ magnitudes of WN-A1 are $10.54 \pm 0.02$  
and $9.43 \pm 0.04$ mag, respectively, while $K$, $L'$ and $M'$ magnitudes of 
WN-A2 are $12.68 \pm 0.02$ , $9.94 \pm 0.02$ and $9.29 \pm 0.03$ mag, respectively. 
The positions of IRS V-1 to V-3 are plotted in Figure \ref{fig1} (D), as well. 

Kraemer et al. (1999) reported four mid-IR (12.5 $\mu$m and 20.6 $\mu$m) sources,
KDJ 1 to 4. As with the above discussion, since relative positions of KDJ 1 and 2
with respect to EN-A1 and A2 agree well with each other, 
we identify EN-A1 and A2 with KDJ 1 and 2, respectively.
In addition, KDJ 3 also corresponds to WN-A1 and A2 collectively. 
Since we have no near-IR counterpart to KDJ 4, the relative position of KDJ 4 with respect to 
KDJ 1 was converted to an absolute position in Table \ref{abs_pos}.

Rengarajan \& Ho (1996) reported three
radio continuum (2 cm) sources. Since R E-2 coincided with IRS V-1, 
we identified R E-2 with EN-A1. As with the position of KDJ 4, the relative positions of 
R E-3 with respect to R E-2 was converted to absolute positions. 
Note that R E-3 was identified with KDJ 4 (Kraemer et al. 1999).
R E-1 has no new near- and mid-IR counterpart and its nature is unknown. Thus,
we do not discuss on this source.

\subsection{Imaging polarimetric observations}\label{3.3}

\begin{figure}
  \begin{center}
    \FigureFile(80mm,50mm){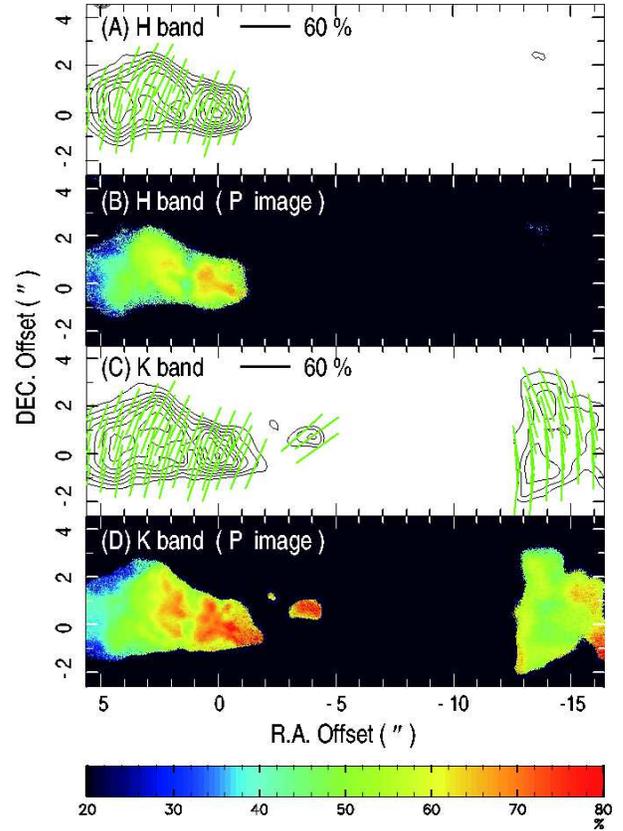}
  \end{center}
  \caption{
    (A) : $H$ band polarization image overlaid with contours of intensity.
    (B) : The degrees of polarization $P$ in the $H$ band.
    (C) : The same as (A) but in the $K$ band.
    (D) : The same as (B) but in the $K$ band. In these all figures,
    the (0$'',0''$) coordinates are the same as in Figure \ref{fig1}.
  }\label{fig2}
\end{figure}

Figure \ref{fig2} shows the various polarization images in the $H$ and $K$ bands,
and table \ref{table1} summarizes the measured degrees of polarization. 
We observed a very large degree of polarizaion for the nebulae in the $H$ and $K$ bands. 
At the peak of EN-A we obtained $\sim 65\%$ in the $H$ band and $\sim 70\%$ 
in the $K$ band. WN-A is polarized at $\sim 60\%$ in the $K$ band. 
These results are consistent with those of Chrysostomou et al. (1994), 
but we obtained a slightly higher degree of polarization than their values. 
We consider that this is due to our high resolution ($0\farcs3$).

  \begin{table}[btph]
      \begin{center}
         \caption{Degrees of polarization in the $H$ and $K$ bands}
	 \label{table1}
      \begin{tabular}{lcc} \hline \hline
	\ \ \ \ \ \ \ \ nebula&this paper&Chrysostomou et al.\\ 
	&&(1994)\\ \hline
	EN-A ( $H$ band )&$\lesssim$ 65 \% &$\lesssim$ 60 \% \\
	EN-A ( $K$ band )&$\lesssim$ 70 \% &$\lesssim$ 60 \% \\
	\\	
	WN-A ( $H$ band )&--- &$\lesssim$ 20 \% \\
	WN-A ( $K$ band )&$\lesssim$ 60 \%&$\lesssim$ 50 \%\\\hline
      \end{tabular}
    \end{center}
  \end{table}
  
\begin{figure}
  \begin{center}
    \FigureFile(80mm,50mm){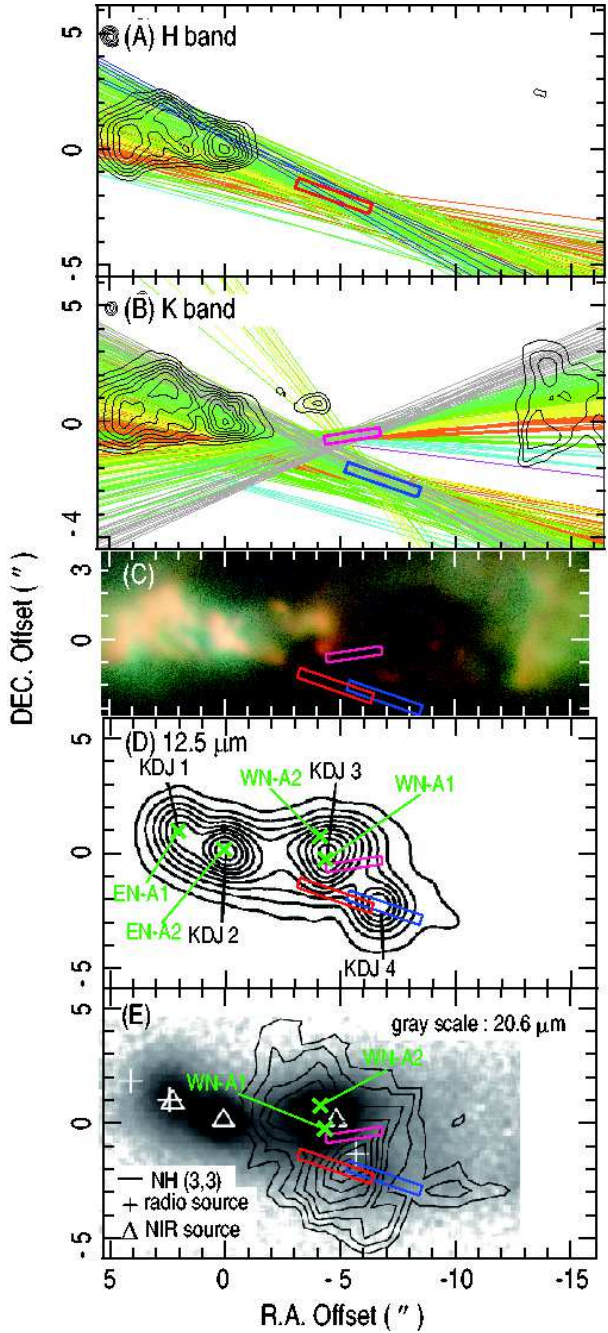}
  \end{center}
  \caption{
    (A): $H$ band polarization image overlaid with contours of total intensity.
    The vectors have been rotated $90^{\circ}$ and extended to indicate the source of 
    the illumination of EN-A. The 1$\sigma$ error box shows this location.
    The color of the vectors are changed for presentation purpose.
    (B): The same as (A) but in the $K$ band.
    (C): The positions of the illuminating sources superposed on the $H$, $K$ and $L'$ 
    composite-image obtained with CIAO.
    (D): Those superposed on the 12.5 $\mu$m contour of Kraemer et al (1999).
    In addition, The positions of WN-A1, WN-A2, EN-A1 and EN-A2 are marked (green crosses of $\times$).
    (E): Those superposed on the NH$_{3}$(3,3) line intensity and the 20 $\mu$m 
    emission of Kraemer et al. (1999).
    The positions of additional features are marked; WN-A1 and A2 (green crosses of $\times$),
    IRS V-1 to V-3 (left to right) at 3.5 $\mu$m and 20 $\mu$m 
    (white triangles of $\triangle$ ; Simon et al. 1985), 
    R E-1 to E-3 (left to right) of 2 cm continuum (white crosses of + 
    ; Rengarajan \& Ho 1996). In these all figures,
    the (0$'',0''$) coordinates are the same as in Figure \ref{fig1}. 
    Note that the position of IRS V-3 is displaced from that in the original
    figure of Kraemer et al. (1999) (see $\S$\ref{3.2} in the text).
  }\label{fig3}
\end{figure}

Figure \ref{fig2}(C) shows that polarization vector pattern in 
EN-A and WN-A differ considerably. In WN-A, the polarization vector patterns
are centrosymmetric, while in EN-A, the vectors are relatively well aligned with
each other, $\theta \sim 160^{\circ}$.
In order to determine the sources of illumination of EN-A and WN-A, 
the polarization vectors were rotated $90^{\circ}$, and extended as shown in 
Figures \ref{fig3}(A) and (B). Figures \ref{fig3}(A) and (B) show the average position 
of all intersections with $1 \sigma$ error box. The errors of the polarization vector patterns 
arisen from the sky noise are quite small. 
The above procedure was based on the seeing size rather than on each pixel.
In order to identify which source illuminates 
the nebulae, the error boxes in Figure \ref{fig3}(A) and (B) are superposed 
on as shown in Figure \ref{fig3}(C), (D) and (E);
NIR composite image (C), MIR contour (D) and  NH$_{3}$(3,3) emission contour (E).

Since both WN-A1 and KDJ 4 lie close to the error boxes, we conclude that
the WN-A is illuminated by WN-A1, and EN-A is by KDJ 4. 
Kraemer et al. (1999) suggested that KDJ 3 and 4 coincided with the 
two NH$_{3}$(3,3) clumps within the positional accuracy.
No additional IR-invisible sources seem to be necessary to interpret our polarization data.

Figures \ref{fig2}(B) and \ref{fig2}(D) show that EN-A1 to EN-A3 and WN-A2 
have more than 50 \% polarization, which implies that these knots are not self-luminous 
in the $H$ and $K$ bands. 
We suggest that these knots seem to be near-infrared single-scattering radiation 
from each relatively high-density region at the walls of the outflows (Nakagawa et al. 1990). 
Nakagawa et al. (1990) explained the exceptionally large (up to $\sim$ 100 \%) degree of 
polarization in the $L'$ band by considering a simple lobe model; the scattering occurred only 
at the surface of the lobe. It has been suggested that there is a close relationship 
between infrared reflection nebulosity (IRN) and CO bipolar outflow (Yamashita et al. 1989). 
Polarimetry toward the central infrared source of CO bipolar outflows 
shows a polarization perpendicular to the outflow direction (Sato et al. 1985). 
Viewed at low spatial resolution, the extension of IRN tends to be consistent with 
that of CO outflow (Tamura et al. 1990). In this NGC 6334 V region, Fischer et al. (1982) 
have shown a CO outflow and Straw, Hyland \& Mcgregor (1989) found evidence that 
this outflow has triggered further star-formation activity via interaction with 
the ambient cloud material. Thus we suggest that EN-A and WN-A are IRNe which are  
the walls of molecular outflows. 
A position-velocity CO map centered on IRS V-3 of Fischer et al. (1982) showed
that two peaks of blueshifted emission were located in the vicinity
and both sides of IRS V-3, which consistents with our results.
Our polarimetric results imply that there are two independent 
outflows in NGC 6334 V region.

\subsection{Natures of the two ''outflow'' sources and other near-IR sources.}

Among the six IR sources, the outflow source KDJ 4 which coincides with 2 cm continuum source R E-3
is identified to be a B2 zero-age main-sequence (ZAMS) star (Rengarajan \& Ho 1996; Kraemer et al. 1999).
Another outflow source WN-A1, however, is not associated with a radio
continuum source (Rengarajan \& Ho 1996), but with (part of) an mid-IR source KDJ 3
and the ammonia clump. Therefore, we suspect this to be a massive embedded YSO with a dense core.

EN-A1 is also associated with a radio coninuum source (R E-2),
thus suggested to be a B-type ZAMS star (Jackson \& Kraemer 1999). 
 However, in \S\ref{3.3}, we mentioned that EN-A1 was a reflection nebula.
Since there is no clear change of near-IR polarization pattern over the
''self-luminous'' star EN-A1, we suggest that 
the scattered light from KDJ 4
is more dominant than the brightness of EN-A1 at near-IR.
By contrast, EN-A2 and WN-A2 (KDJ 2 and part of KDJ 3, respectively) are not associated with radio 
coninuum sources (Rengarajan \& Ho 1996) and not temperature peaks (Kraemer et al. 1999)
, therefore are knots of the near-IR reflection nebula.
Similarly, EN-A3 seems to be a simply peak of the near-IR reflection nebula.  

Therefore, our observations, combined with previous mid-IR and radio observations, 
have revealed the presence of at least three clustered YSOs in the NGC 6334 V,
of which two YSOs are probably powering massive outflows.
Further high spatial resolution CO observations 
should be undertaken to reveal the kinematics of these outflows.

\section{Conclusion}\label{concl}
We conducted Subaru/CIAO direct and polarimetric imaging of the NGC 6334 V nebulae;
Eastern Nebula A (EN-A) and Western Nebula A (WN-A), 
with a high resolution of 0$\farcs$3. We detected two red near-IR sources (WN-A1 and A2)
near the position of mid-IR source KDJ 3. We found that EN-A and WN-A 
are illuminated by the different infrared sources of KDJ 4 and WN-A1, respectively.
Combined with previous mid-IR and radio observations,
we suggest that the outflow sources of KDJ 4 and WN-A1 are a B-type ZAMS star
and a massive embedded YSO with a dense core, respectively.
In addition, fine structures of both nebulae were detected; in particular,
we found that the western nebula has a multi-shell structure.
The high levels of polarization of EN-A2 and WN-A2 in the $K$ band suggest that
these are not self-luminous, but rather knots of the reflection nebulae.

\bigskip

We are grateful to Misato Fukagawa for several useful comments on an earlier version of this paper.
We also thank an anonymous referee for providing
many useful comments that lead to a significantly improved paper.
This research was partly supported by MEXT Japan,
Grant-in-Aid for Scientific Research
on Priority Areas, "Development of
Extra-solar Planetary Science".

\end{document}